\documentclass{appolb}

\usepackage[T1]{fontenc}

\usepackage{amssymb}
\usepackage{graphicx}
\usepackage{amsmath}
\usepackage{subfigure}
\usepackage{bbold}
\usepackage{yfonts}
\usepackage{placeins}
\usepackage{bm} 
\usepackage{nicefrac}
\usepackage{slashed}

\newcommand{\bea}{\begin{eqnarray}}
\newcommand{\eea}{\end{eqnarray}}
\newcommand{\bel}[1]{\begin{eqnarray}\label{#1}}
\newcommand{\eel}{\end{eqnarray}}

\def\LB{\left(}
\def\RB{\right)}

\def\spin{\,\textgoth{s:}}

\def\Lg{\,L}

\newcommand{\bv}{{\boldsymbol b}} 
\newcommand{\ev}{{\boldsymbol e}} 

\newcommand{\xv}{{\boldsymbol x}}

\newcommand{\pv}{{\boldsymbol p}}

\newcommand{\Pv}{{\boldsymbol P}}

\newcommand{\ppiv}{{\boldsymbol \pi}}
\newcommand{\sigv}{{\boldsymbol \sigma}}

\def\pmU{p^\mu}

\newcommand{\p}{\partial}

\newcommand{\f}[2]{\frac{#1}{#2}}

\newcommand{\tr}{{\rm tr}}

\newcommand{\trt}{{\rm tr_2}}
\newcommand{\trf}{{\rm tr_4}}

\def\bmL{\beta_\mu}

\def\omnL{\omega_{\mu\nu}}

\def\epsUabgd{\epsilon^{\alpha \beta \gamma \delta}}

\def\epsLmnab{\epsilon_{\mu\nu\alpha\beta}}

\def\g5{\gamma_5}

\def\TmnU{T^{\mu\nu}}
\def\TnmU{T^{\nu\mu}}

\def\oTmnU{{\widehat T}^{\mu\nu}}

\def\SmnU{{\Sigma}^{\mu\nu}}

\def\S0iU{{\Sigma}^{0i}}

\def\SlmnU{S^{\lambda, \mu\nu}}
\def\SmlnU{S^{\mu, \lambda\nu}}

\def\oSmlnU{{{\widehat S}^{\mu, \lambda\nu}}}

\def\wj{{\widehat j}}

\def\n0{n_{(0)}}
\def\e0{\varepsilon_{(0)}}
\def\P0{P_{(0)}}

\def\TmnU{T^{\mu\nu}}                      % energy-momentum tensor

\def\rhoLEQ{{\widehat{\rho}}_{\rm \small LEQ}}

\def\fplusrsxp{f^+_{rs}(x,p)}

\def\fminusrsxp{f^-_{rs}(x,p)}

\def\ubarrp{{\bar u}_r(p)}

\def\usp{u_s(p)}

\def\vbarsp{{\bar v}_s(p)}

\def\vrp{v_r(p)}

\def\Wpmxk{{\cal W}^{\pm}(x,k)}
\def\Wxk{{\cal W}(x,k)}

\def\Weqpxk{{\cal W}^{+}_{\rm eq}(x,k)}

\def\Fpmxk{{\cal F}^{\pm}(x,k)}

\def\Ppmxk{{\cal P}^{\pm}(x,k)}

\def\spin{\,\textgoth{s:}}

\def\Lg{\,L}

\begin{document}
\title{Hydrodynamics with spin --- recent developments%
\thanks{Presented at the Epiphany 2019 Conference, 8-11 January 2019, Krak\'ow, Poland }%
}
\author{Wojciech Florkowski
\address{Marian Smoluchowski Institute of Physics, Jagiellonian University \\
ul. \L ojasiewicza 11, 30-348 Krak\'ow, Poland
}
\\
}
\maketitle
\begin{abstract}
Recent theoretical developments in hydrodynamics of particles with spin \nicefrac{1}{2} are briefly reviewed.
\end{abstract}
  
\section{Introduction}
In the relativistic non-central heavy-ion collisions, large amount of the initial orbital angular momentum can be transferred into produced systems. A noticeable part of such an angular momentum can be further relocated from the orbital part to the spin component. The latter can be eventually displayed in the spin polarization of emitted particles such as $\Lambda$ and  $\bar{\Lambda}$ hyperons~\cite{Voloshin:2004ha,Liang:2004ph,Betz:2007kg,Gao:2007bc,Voloshin:2017kqp}. The spin polarization of  both $\Lambda$'s and  $\bar{\Lambda}$'s has been indeed measured by the STAR Collaboration ~\cite{STAR:2017ckg, Adam:2018ivw}. The experimental result shows global, out-of-plane spin polarization, which suggests  connections to the Einstein~--~de~Haas and Barnett effects~\cite{dehaas:1915,RevModPhys.7.129}. 

 The appearance of global polarization has been successfully explained by the hydrodynamic models~\cite{Karpenko:2016jyx} that directly connect spin polarization effects with the thermal vorticity. The latter is defined  by the expression $\varpi_{\mu \nu} = -\frac{1}{2} (\partial_\mu \beta_\nu-\partial_\nu \beta_\mu)$, where $\beta_\mu$ is the ratio of the flow velocity $u_\mu$ to local temperature $T$, $\beta_\mu = u_\mu/T$~\cite{Becattini:2007nd,Becattini:2009wh,Becattini:2016gvu}. There remain, however, questions concerning description of the longitudinal polarization, since the theoretically predicted longitudinal polarization of $\Lambda$'s~\cite{Becattini:2017gcx} has an opposite sign of the dependence on the azimuthal angle of the emitted particles, compared to the experimentally found results~\cite{Niida:2018hfw}. 
 
 On the general thermodynamic grounds \cite{Becattini:2018duy} one can expect that the spin polarization effects, quantified by the tensor $\omega_{\mu\nu}$ (dubbed below the spin polarization tensor), can be independent of the thermal vorticity  $\varpi_{\mu \nu}$. This idea was proposed first in Ref.~\cite{Florkowski:2017ruc} and developed in Refs.~\cite{Florkowski:2017dyn,Florkowski:2018myy,Florkowski:2018ahw} (for a recent review see Ref.~\cite{Florkowski:2018fap} and for related works see Refs.~\cite{Sun:2018bjl,Weickgenannt:2019dks,Gao:2019znl}).  In the approach proposed in~\cite{Florkowski:2017ruc}, which forms the basis for relativistic perfect-fluid hydrodynamics of particles with spin \nicefrac{1}{2}, the space-time evolution of the polarized fluid is determined by the conservation laws including the conservation of total angular momentum (we note that for particles with spin, the latter has a non-trivial form). 
  
Inclusion of the spin polarization into the hydrodynamic framework is appealing, as the present works say little about the changes of the spin polarization during the heavy-ion collision process. As the fluid dynamics has become now the basic ingredient of heavy-ion  models (for recent developments within relativistic hydrodynamics see \cite{Florkowski:2017olj,Romatschke:2017ejr}), it is even more interesting to have spin effects included into the hydrodynamic picture of heavy-ion collisions. So far, relatively little work has been done in this direction, although the studies of fluids with spin have a rather long history that started in 1940s \cite{Weyssenhoff:1947iua,Bohm:1958,Halbwachs:1960aa}. Below, we outline recent developments done in this field. 

\section{Generalized local equilibrium}

The concept of perfect-fluid hydrodynamics with spin is based on the idea of a generalized thermodynamic equilibrium which is described, in addition to the standard hydrodynamic quantities such as temperature $T(x)$, flow four-vector $u^\mu(x)$, and chemical potential $\mu(x) = \xi(x) T(x)$, by the spin polarization tensor $\omega_{\mu\nu}(x)$~\cite{Becattini:2018duy}. One uses those quantities to construct the density operator $\rhoLEQ$ and to obtain the expectation values of the energy-momentum tensor $\TmnU$, the spin tensor $\SmlnU$, and the baryon current $j^{\mu}$ from the corresponding operators:
\bea
\TmnU =  \tr \left(\rhoLEQ \, \oTmnU \right), \,\,
\SmlnU = \tr \left(\rhoLEQ \,  \oSmlnU \right), \,\,
j^{\mu} = \tr \left(\rhoLEQ \, \wj^{\mu}\right).
\label{eq:rh0}
\eea
Thus, we can write
\bea
\TmnU = \TmnU [\beta,\omega,\xi], \,\,
\SmlnU = \SmlnU [\beta,\omega,\xi], \,\,
j^\mu = j^\mu [\beta,\omega,\xi].
\label{eq:TSj}
\eea
In local equilibrium with dissipation effects neglected, one can assume that the density operator is constant, which leads to the following equations:
\bea
\p_\mu \TmnU = 0, \quad
\p_\lambda \SlmnU = \TnmU -\TmnU, \quad
\p_\mu  j^\mu =0.
\label{eq:h1}
\eea
These are 11 equations for 11 unknown functions (temperature, three independent components of the fluid four-velocity, chemical potential, and 6 independent components of the tensor $\omega_{\mu\nu}$ which becomes now a new hydrodynamic variable). We note that in general only the total angular momentum is conserved, which leads to the middle equation in (\ref{eq:h1}). 

%%%%%%%%%%%%%%%%%%%%%%%%%%%%%%%%%%%
\section{Spin-dependent phase-space distribution functions}

The general framework defined in the previous section illustrates the concept of the hydrodynamics with spin but,  in practice, we need more explicit forms of the energy-momentum and spin tensors. To obtain such forms it is useful to introduce first the spin-dependent  phase-space distributions functions~\cite{Becattini:2013fla},
\bea
\left[ f^+(x,p) \right]_{rs} &=& \fplusrsxp =  \ubarrp X^+ \usp, \nonumber  \\
\left[ f^-(x,p) \right]_{rs} &=& \fminusrsxp = - \vbarsp X^- \vrp. 
  \label{eq:frs}
 \eea
Here $r,s=1,2$ are spin indices, and $u$ and $v$ are Dirac bispinors (with $x = (t, \xv)$ and $p = (E_p = p^0, \pv)$, where $p^0=E_p=\sqrt{m^2+\pv^2}$\,). The matrices $X^\pm$ are defined by the expressions
\bea
X^{\pm} =  \exp\left[\pm \xi(x) - \bmL(x) \pmU \right] M^\pm,  \quad
M^\pm = \exp\left[ \pm \f{1}{2} \omnL(x)  \SmnU \right],
\eea
where $\SmnU$ is the Dirac spin operator.  It is convenient to use the parametrization of the spin tensor in terms of the electric- and magnetic-like three-vectors
\bel{omeb}
\omnL= 
\begin{bmatrix}
	0       &  e^1 & e^2 & e^3 \\
	-e^1  &  0    & -b^3 & b^2 \\
	-e^2  &  b^3 & 0 & -b^1 \\
	-e^3  & -b^2 & b^1 & 0
\end{bmatrix}.
\label{eq:eb}
\eel
Herein, we restrict our considerations to the case where $\omega_{\mu\nu}$ is small and keep only leading terms in $\omega_{\mu\nu}$.

The phase-space distributions can be written in the matrix form employing the Pauli matrices $\sigv = (\sigma^1, \sigma^2, \sigma^3)$, namely
\bel{fpm}
f^\pm(x,p) =  e^{\pm \xi - p \cdot \beta} \left[ 1  - \f{1}{2}  \, \Pv \cdot \sigv \right],
\label{eq:f}
\eel
where
\bea
\Pv &=& \f{1}{m} \left[  E_p \, \bv - \pv \times \ev - \f{\pv \cdot \bv}{E_p + m} \pv \right]  = \bv_\ast .
\quad 
\label{eq:Pv}
\eea
Here $E_p = \sqrt{m^2+\pv^2}$ and the asterisk denotes the value of the $\bv$ field in the particle rest frame (PRF).
From (\ref{eq:Pv}) we obtain the average spin polarization per particle~\cite{Leader:2001}
\bea
\left\langle \Pv(x,p) \right\rangle =  \f{1}{2} \f{ \trt \left[ (f^+ + f^-) \sigv\right]  }{\trt \left[ f^+ + f^- \right] }  = -\f{1}{4} \Pv ,
\label{eq:vecP}
\eea
where $\trt$ denotes the trace over spin indices. Hence, the knowledge of the spin-dependent phase-space distributions allows us to obtain the information about the spin polarization of particles.

%%%%%%%%%%%%%%%%%%%%%%%%%%%%%%%%%%%%%%%%%
\section{FFJS hydrodynamic model for particles with spin \nicefrac{1}{2}}

The hydrodynamic model formulated in~\cite{Florkowski:2017ruc} uses the phase-space distributions introduced above to define the baryon current $j^{\mu}$,  the energy-momentum tensor $T^{\mu\nu}$, and the spin tensor $S^{\lambda,\mu\nu}$ according to the following prescriptions (with $\trf$ denoting now the trace over spinor indices):
\bea
j^{\mu} = \int \frac{d^3p}{2 (2\pi)^3 E_p} p^\mu \left[ \trf (X^+) +  \trf (X^-) \right],
\eea
\bea
T^{\mu\nu} = \int \frac{d^3p}{2 (2\pi)^3 E_p} p^\mu p^\nu \left[ \trf (X^+) +  \trf (X^-) \right],
\eea
\bea
S^{\lambda,\mu\nu} = \int \frac{d^3p}{2 (2\pi)^3 E_p} p^\lambda  \trf \left[(X^+ - X^-) \Sigma^{\mu\nu} \right].
\eea
These forms, while used in the general scheme defined by Eqs.~(\ref{eq:h1}), allow us for the construction of the hydrodynamic framework with spin degrees of freedom, which turns out to be thermodynamically consistent and entropy conserving. In spite of such appealing features, further studies showed that one has to switch to other forms of those tensors. We explain this development in the next two section.

%%%%%%%%%%%%%%%%%%%%%%%%%%%%%%%%%%%%%%%%%
\section{Semiclassical kinetic equation}

The equilibrium phase-space distribution functions discussed above can be used to obtain the corresponding equilibrium Wigner functions. For example, one can find~\cite{DeGroot:1980dk}
\bea
\Weqpxk = \frac{1}{2} \sum_{r,s=1}^2 \int \frac{d^3p}{(2\pi)^3 E_p}\,
\delta^{(4)}(k-p) u^r(p) {\bar u}^s(p) f^+_{rs}(x,p),
\nonumber
\eea
and a similar expression can be introduced for antiparticles (here $k$ is the four-momentum that in general can be off-shell). Any Wigner functions can be furthermore expressed as combinations of the generators of the Clifford algebra~\cite{Elze:1986hq,Elze:1986qd,Vasak:1987um,Zhuang:1995pd,Florkowski:1995ei,Gao:2017gfq}
\bea
&& \Wpmxk =  \f{1}{4} \left[ \Fpmxk + i \gamma_5 \Ppmxk + \gamma^\mu {\cal V}^\pm_{ \mu}(x,k) \right.  \nonumber \\
&& \hspace{3cm} \left. + \gamma_5 \gamma^\mu {\cal A}^\pm_{ \mu}(x,k)
+ \SmnU {\cal S}^\pm_{ \mu \nu}(x,k) \right],   \label{eq:W}
\eea
which is a very useful tool for studying a semiclassical limit of the quantum kinetic equation. Such an equation has the form
\bel{eq:eqforWC}
\left(\gamma_\mu K^\mu - m \right) {\cal W}(x,k) = C[{\cal W}(x,k)],
\label{eq:kineq}
\eel
where  $K^\mu$ is the operator defined by the expression
\bel{eq:K}
K^\mu = k^\mu + \frac{i \hbar}{2} \,\p^\mu \label{eq:Kmu}
\eel
and $C[{\cal W}(x,k)] $ is the collision term. In global or local equilibrium, the collision term vanishes and
one can study only the left-hand side of (\ref{eq:kineq}) that should be equal to zero in this situation.

At this point it is important to distinguish between the global and local equilibrium. In the case of global equilibrium the Wigner function ${\cal W}(x,k)$ exactly satisfies the equation
\bel{eq:eqforW}
\left(\gamma_\mu K^\mu - m \right) {\cal W}(x,k) = 0.
\label{eq:kineq0}
\eel
If the leading order (of the expansion of $ {\cal W}$ in $\hbar$) is taken as  ${\cal W}_{\rm eq}$, one finds (from the leading and next-to-leading terms in $\hbar$) that $\mu/T$~=~const., $\omega_{\mu\nu}$~=~const., and $\partial_\mu \beta_\nu - \partial_\nu \beta_\mu = 0$. The last equation is known as the Killing equation that (in the flat space-time) has a solution $\beta_\mu = b^0_\mu + \omega^0_{\mu \nu} x^\nu$ with  $b^0_\mu$~=~const. and $\omega^0_{\mu \nu} = - \omega^0_{\nu \mu} $~=~const. Interestingly, the tensors $\omega_{\mu\nu}$ and $\omega^0_{\mu\nu}$ might be different. We define such a case as extended global equilibrium, while the name global equilibrium is restricted to the situation where  $\omega_{\mu\nu}=\omega^0_{\mu\nu}$.

In the case of local equilibrium, one assumes that only specific moments of Eq.~(\ref{eq:kineq0}) vanish. This point has been discussed in more detail in Ref.~\cite{Florkowski:2018ahw}, where it is shown that this procedure leads to the following equations
\bea
\p_\mu j^\mu_{\rm GLW}(x)  = 0, \quad \p_\alpha T^{\alpha\beta}_{\rm GLW}(x) = 0, \quad
\p_\lambda S^{\lambda , \mu \nu }_{\rm GLW}(x) = 0.
\label{eq:GLWhydro}
\eea
This form is again consistent with the general scheme of the hydrodynamics with spin, however, the forms of the tensors appearing in Eqs.~(\ref{eq:GLWhydro}) are different from those used in the FFJS model. As a matter of fact,  these forms agree with the expressions used by de Groot, van Leeuwen, and van Weert in Ref.~\cite{DeGroot:1980dk}. We note that the numerical solutions of Eqs.~(\ref{eq:GLWhydro}) has been obtained recently in Ref.~\cite{Florkowski:2019qdp}.

%%%%%%%%%%%%%%%%%%%%%%%%%%
\section{Pseudo-gauge transformations}

The kinetic equation (\ref{eq:kineq}) always mixes terms appearing in the two neighbouring orders of the expansion in $\hbar$. In the leading order, it produces algebraic relations that are automatically fulfilled by ${\cal W}_{\rm eq}$. Only in the next-to-leading order it gives differential equations that can be interpreted as the truly kinetic equations. Besides such equations, in the next-to-leading order one gets also the expressions that define corrections to the pseudoscalar, vector, and tensor constributions to ${\cal W}$,
\bel{eq:rP1aeq}
{\cal P}^{(1)}  = -\frac{1}{2m} \, \p^\mu  {\cal A}_{\rm eq, \mu} \,,
\label{eq:P1}
\eel
\bel{eq:rV1aeq}
{\cal V}^{(1)}_\mu &=& -\frac{1}{2m} \p^\nu {\cal S}_{\rm eq, \nu \mu} \,,
\label{eq:V1}
\eel
\bel{eq:rS1aeq}
{\cal S}_{\mu \nu}^{(1)} = \frac{1}{2m} \left(\p_\mu {\cal V}_{\rm eq,\nu} - \p_\nu {\cal V}_{\rm eq,\mu} \right) .
\label{eq:S1}
\eel
These corrections help us to understand the difference between the GLW forms and the canonical forms of the energy-momentum and spin tensors. For example, in the case of the energy-momentum tensor, we find
\bel{eq:tmunu1}
T^{\mu\nu}_{\rm GLW}(x)=\frac{1}{m}\trf \int d^4k \, k^{\mu }\,k^{\nu }\Wxk=\frac{1}{m} \int d^4k \, k^{\mu }\,k^{\nu } {\cal F}(x,k) \label{eq:TGLW}
\eel
and
\bel{eq:tmunu1can1}
T^{\mu\nu}_{\rm can}(x)= \int d^4k \,k^{\nu } {\cal V}^\mu(x,k) \label{eq:Tcan}.
\eel
We thus see that quantum corrections induce asymmetry of the canonical tensor, $T^{\mu\nu}_{\rm can}(x) \neq T^{\nu\mu}_{\rm can}(x)$. Similarly, we find differences between $S^{\lambda , \mu \nu }_{\rm GLW}$ and $S^{\lambda , \mu \nu }_{\rm can}$. 

It is very interesting to observe that if we introduce a tensor $\Phi_{\rm can}^{\lambda, \mu\nu}$ defined by the relation
\bel{Phi}
\Phi_{\rm can}^{\lambda, \mu\nu} 
\equiv S^{\mu , \lambda \nu }_{\rm GLW}
-S^{\nu , \lambda \mu }_{\rm GLW} \label{eq:Phi}
\eel
we can write
\bel{psg1GLW}
S^{\lambda , \mu \nu }_{\rm can}= S^{\lambda , \mu \nu }_{\rm GLW} -\Phi_{\rm can}^{\lambda, \mu\nu}
\eel
and
\bel{psg2GLW}
T^{\mu\nu}_{\rm can} = T^{\mu\nu}_{\rm GLW} + \frac{1}{2} \p_\lambda \left(
\Phi_{\rm can}^{\lambda, \mu\nu}
+\Phi_{\rm can}^{\mu, \nu \lambda} 
+ \Phi_{\rm can}^{\nu, \mu \lambda} \right) .
\eel
Hence, the canonical and GLW frameworks are connected by a pseudo-gauge transformation. 

The conclusion that can be drawn from the last two sections is that the GLW framework offers a well defined scheme to define hydrodynamics with spin: it is well based on the kinetic theory and has a natural connection to the canonical formalism. The pseudo-gauge transformation connecting the canonical expressions with the GLW ones is similar to the Belinfante construction but it does not eliminate the spin tensor which can be used to describe spin degrees of freedom. 

%%%%%%%%%%%%%%%%%%%%%%%%%% 
\section{Pauli-Luba\'nski four-vector}

It is also interesting to demonstrate that in our approach one can introduce the Pauli-Luba\'nski (PL) four-vector (strictly speaking its classical phase-space density) which describes the spin polarization in the way consistent with that based on the phase-space densities defined in Sec.~3. We introduce first the quantity 
\bel{PL10}
E_p \f{d \Delta \Pi_\mu(x,p)}{d^3p}  = -\f{1}{2} \epsLmnab \, \Delta \Sigma_\lambda(x) \, 
E_p \f{d J^{\lambda, \nu\alpha}(x,p)}{d^3p}
\f{p^\beta}{m} .
\eel
One can check that only the spin-part contributes here and the results are the same for the canonical and GLW versions.
By dividing $E_p \f{d \Delta \Pi_\mu(x,p)}{d^3p}$ by the total density of particles and antiparticles, we find
\bel{PL4}
\pi_\mu(x,p) \equiv \f{\Delta \Pi_\mu(x,p)}{\Delta {\cal N}(x,p)} &=& 
-\f{\hbar}{4 m} \, {\tilde \omega}_{\mu \beta} \, p^\beta ,
\eel
where $ {\tilde \omega}_{\mu \nu} $ is the dual polarization tensor. In PRF this result is reduced to
\bel{pi0PRF}
\pi^0_\ast   = 0, \qquad {\ppiv}_\ast   =   - \f{\hbar }{4}  \Pv,
\eel
which is consistent with (\ref{eq:vecP}).

%%%%%%%%%%%%%%%%%%%%%%%%%%%%%%%%%%%%%%%
\section{Classical treatment of spin}

Finally, let us discuss very recent results which make use of the classical concept of spin. We introduce the intrinsic angular momentum following the works by Mathisson~\cite{Mathisson:1937zz}
\bea
s^{\alpha\beta} = \f{1}{m} \epsUabgd p_\gamma s_\delta.
\label{eq:sab}
\eea
Here $s^\mu$ is the spin four-vector that is orthogonal to the four-momentum
\bea
s \cdot p = 0.
\label{eq:sp}
\eea
A straightforward generalization of the phase-space distribution function $f(x,p)$ is a spin dependent distribution $f(x,p,s)$. Accordingly, we have to consider an extended phase space, whose element is $dSdP$ with
\bea
\int dS \ldots = \frac{m}{\pi \spin}  \, \int d^4s \, \delta(s \cdot s + \spin^2) \, \delta(p \cdot s) \ldots \,,
\label{eq:dS}
\eea
where the length squared of the spin vector in PRF is taken to be the value of the Casimir operator for spin \nicefrac{1}{2}, 
\bea
\spin^2 = \f{1}{2} \left( 1+ \f{1}{2}  \right) = \f{3}{4}.
\label{eq:spin2}
\eea
The overall normalization $m/(\pi \spin)$ in $dS$ is also taken in such a way as to describe spin \nicefrac{1}{2} particles,
\bea
\int dS = \f{m}{\pi \spin}  \int \, d^4s \, \delta(s \cdot s + \spin^2) \, \delta(p \cdot s) = 2.
\label{eq:dSint}
\eea

The equilibrium distribution functions for particles and antiparticles can be written in the exponential form~\cite{Weert:1970aa}
\bea
f^\pm_{\rm eq}(x,p,s) = \exp\LB - p \cdot \beta(x) \pm \xi(x) + \f{1}{2}  \omega_{\alpha \beta}(x) s^{\alpha\beta} \RB
\label{eq:fpm-spin}
\eea
and the basic currents take natural forms:
\bea
j^\mu_{\rm eq} = \int dP   \int dS \, \, p^\mu \, \left[f^+_{\rm eq}(x,p,s)-f^-_{\rm eq}(x,p,s) \right],
\label{eq:Neq-sp0}
\eea
\bea
T^{\mu \nu}_{\rm eq} &=& \int dP   \int dS \, \, p^\mu p^\nu \, \left[f^+_{\rm eq}(x,p,s) 
+ f^-_{\rm eq}(x,p,s) \right] \label{eq:Neq-sp01},
\eea
\bea
\hspace{-0.75cm}S^{\lambda, \mu\nu}_{\rm eq} &=& \int dP   \int dS \, \, p^\lambda \, s^{\mu \nu} 
\left[f^+_{\rm eq}(x,p,s) + f^-_{\rm eq}(x,p,s) \right] .
\label{eq:Seq-sp01}
\eea

An intriguing observation one can make is that in the leading order of the expansion in $\omega_{\mu\nu}$ we obtain the formalism that agrees with that based on the quantum description of spin (in the GLW version). 

On the other hand, for arbitrary values of $\omega_{\mu\nu}$, the PL four-vector can be expressed by the expression
\bea
\pi_\mu = - \spin \, \f{{\tilde \omega_{\mu \beta}}}{P} \, \f{p^\beta}{m} \, \Lg(P  \spin),
\label{PL1}
\eea
where $\Lg(x)$ is the Langevin function defined by the formula
\bea 
\Lg(x) =  \coth(x) - \f{1}{x}.
\label{eq:Lg}
\eea 

For small and large values of $P$ we obtain two important results:
\bea 
\ppiv_* = -  \spin\,\f{\Pv}{P}, \quad
|\ppiv_*| = \spin = \sqrt{\f{3}{4}}, \quad
\hbox{if} \quad P \gg 1
\label{eq:PLlarge}
\eea 
and
\bea 
\ppiv_* = - \spin^2\,\f{\Pv}{3}, \quad
|\ppiv_*| = \spin^2\, \f{P}{3} = \f{P}{4}, \quad
\hbox{if} \quad P \ll 1.
\label{eq:PLsmall}
\eea 

The results presented in this section give us hints about the behavior of systems with large spin polarization. Clearly, they become anisotropic, since large spin polarization introduces a privileged direction in space and pressure cannot be isotropic. This suggests that the methods of anisotropic hydrodynamics \cite{Florkowski:2010cf,Martinez:2010sc} could be applied to further study such systems. 

%%%%%%%%%%%%%%%%%%%%%%%%%% 
\section{Conclusions}

The results presented in this contribution describe dynamics of a~perfect fluid consisting of particles with spin \nicefrac{1}{2}. Several works reported herein clarify the use of different forms of the energy-momentum and spin tensors and conclude that the GLW expressions are the most appropriate --- their use follows from the kinetic-theory analysis and they are related with the canonical expressions (obtained by the Noether theorem) through a pseudo-gauge transformation. 

The main challenge for the next developments of the hydrodynamics with spin is the proper inclusion of dissipation (which includes, in particular, a calculation of  kinetic coefficients related to spin observables). First steps in this direction have been made, for example, in Ref.~\cite{Hattori:2019lfp}. In the closest future, it would be interesting to examine more closely the relation of the results presented in Ref.~\cite{Hattori:2019lfp} to the formalism discussed in this text. It is also mandatory to study in more detail the relation between spin polarization and thermal vorticity. An effect describing convergence of the spin polarization tensor to the thermal vorticity should be included in the complete formalism of viscous hydrodynamics with spin.

\medskip
This work was supported in part by the Polish National Science Center Grant  No. 2016/23/B/ST2/00717.

\end{document}